# Migration of Giant Gaseous Clumps and Structure of the Outer Solar System

V. V. Emel'yanenko

Institute of Astronomy, Russian Academy of Sciences, Moscow, 109017 Russia

*e-mail:* vvemel@inasan.ru

**Abstract.** New data on the distribution of distant trans-Neptunian objects and on the properties of comets indicate the importance of dynamical processes in the outer part of the protoplanetary disk in the formation of the observed structure of the Solar system. In this paper, we examined the possible action of giant gaseous clumps, resulting from gravitational instability and fragmentation of circumstellar disks, on the orbital distribution of the population of small bodies in the outer Solar system. Basically, we studied those features of migration and gravitational interaction of giant clumps that were found previously by Vorobyov and Elbakyan (2018). Our modeling showed that the main features of the distribution of small bodies resulting from the gravitational influence of giant clumps are consistent with the observed orbital distribution of distant trans-Neptunian objects. The studied dynamical process associated with a single giant clump is very short-time event (no more than several tens of thousands of years). The main factor affecting the orbital distribution of small bodies is close encounters with giant clumps. A significant part of small bodies (comets) is very quickly transferred to distant orbits with large eccentricities, which allows them to avoid mutual collisions.

## INTRODUCTION

The recent discovery of distant trans-Neptunian objects moving in orbits with semimajor axes $a > 150$ AU, gave new and rather unexpected information about the structure of the outer part of the Solar system. In the distribution of the angular orbital elements of these objects, an unusual grouping was found near certain values. Initially, Trujillo and Sheppard (2014) suggested that there is a concentration of arguments of perihelion $\omega$ near the value $\omega = 0°$. This value corresponds to one of the centers of the libration zones in the Lidov–Kozai mechanism of secular perturbations; therefore, an assumption was made about the existence of a distant planet producing this effect. Later, it was stated (Batygin, Brown, 2016) that a grouping of longitudes of perihelion $\pi$ and longitudes of ascending nodes $\Omega$ is rather the cause and this effect is associated with the combined action of orbital and secular resonances (Batygin, Morbidelli, 2017). As a result of detailed celestial-mechanical studies, it was shown that the observed features in the orbital distribution for distant trans-Neptunian objects could be produced by a planet with mass of ~10 Earth masses moving in an orbit with semimajor axis of ~400–800 au, eccentricity of ~0.2–0.5 and inclination of ~15–25° (Batygin, Brown, 2016; Batygin et al., 2019).

Although the dynamical picture looks quite convincing (Batygin, Morbidelli, 2017), the question of the actual existence of the ninth planet of the Solar system remains open. Despite intensive searches, the planet has not yet been discovered. The explanation of the formation of such a massive and distant planet is also of great difficulty (Batygin et al., 2019). Therefore, there is a need to consider other possible scenarios for the formation of the observed structure of the distant trans-Neptune region.

New questions in the theory of the formation of remote regions of the solar system were posed by the results of the Rosetta space mission to Comet 67P / Churyumov-Gerasimenko. The origin of comets has always been associated with the outer Solar system. Since the time of both physical and dynamical life of comets in near-Earth space is very short compared with the age of the Solar system, it is natural to assume that the comets spent most of their time in remote regions of the Solar system. In the last decade, the point of view that the formation of cometary populations occurred during the migration of planets described by the Nice model (Levison et al., 2011) has become most widespread. In the work (Brasser, Morbidelli, 2013) it was emphasized that within the framework of this model, the common origin of the trans-Neptunian objects and the objects of the Oort cloud from a single population of planetesimals located initially beyond the orbit of Neptune is naturally explained. However, new data on the physical properties of Comet 67P/Churyumov-Gerasimenko obtained during the Rosetta space mission are in poor agreement with the Nice model in terms of the long (~0.5 Myr) existence of the initial planetesimal disk beyond the orbits of giant planets in the region of 15 -30 au (Davidsson et al., 2016). In the works (Morbidelli, Rickman, 2015; Rickman et al., 2015), it was shown that cometary objects of kilometer sizes had to experience numerous mutual collisions during the considered period of time. Therefore, in the Nice model, cometary nuclei are not the original planetesimals, but are shock fragments arising from the destruction of bodies that were initially much larger. But the general conclusion made in (Davidsson et al., 2016) based on data analysis for Comet 67P/Churyumov-Gerasimenko in comparison with numerous ground-based observations and the results of other space missions, as well as laboratory experiments and numerical calculations, is that that comets are the primordial objects that have not experienced large shock and thermal changes since the formation. Furthermore, in (Fulle et al., 2016; Mannel et al., 2016; Fulle, Blum, 2017) it was stressed that the existence of very fluffy particles , also called fractal, in Comet 67P/Churyumov-Gerasimenko implies possible collisions only with relative velocities less than 1 m/s to preserve such structures.

Thus, the data of the Rosetta space mission also indicate the need to consider new models in which the accretion of distant objects occurred at extremely low velocities, and the transition to orbits with high eccentricities, typical for objects of the trans-Neptune scattered disk and Oort cloud (highly probable sources of observed comets) occurred much earlier than in the Nice model.

Gravitational instability and fragmentation of circumstellar disks has long been regarded as a possible mechanism for the rapid formation of giant planets (Boss, 1997). Currently, there is a fairly well-established point of view that the conditions for the fragmentation of the circumstellar disk are satisfied in its outer part, at distances of more than 50–100 AU from the star (e.g., Johnson, Gammie, 2003; Rafikov, 2005). The detection of exoplanets at wide orbits is an observational support for these theoretical estimates. The minimum initial mass of self-gravitating gaseous clumps arising from disk fragmentation is ~$10m_J$, where $m_J$ is the mass of Jupiter (e.g., Stamatellos and Whitworth, 2009), and the initial radii of such formations, considered in simulations of disk fragmentation, range from from several astronomical units (Boss, 2011; Galvagni, Mayer, 2014) to several tens of astronomical units (Vorobyov, 2013). Recently, it has been shown that the resulting giant clumps participate in a complex dynamical interaction with the disk, which leads, in particular, to the migration of clumps (Mayer et al., 2002; Vorobyov, Basu, 2005; Nayakshin, 2010; Baruteau et al., 2011; Zhu et al., 2012; Stamatellos, 2015; Vorobyov, Elbakyan, 2018). Moreover, migrating clumps can experience close encounters with each other, which often leads to the ejection of objects into hyperbolic orbits (Terquem, Papalozou, 2002; Vorobyov, Elbakyan, 2018).

FEATURES OF THE ORBITAL DISTRIBUTION OF DISTANT TRANS-NEPTUNIAN OBJECTS

The distribution of the angular orbital elements of distant trans-Neptunian objects caused much discussion and became the basis for the assumption of the existence of the ninth planet. Here we consider the distribution of angular elements for objects with perihelion distances $q > 40$ au, in contrast to the works (Trujillo, Sheppard, 2014; Batygin, Brown, 2016), in which the orbits of objects with $q > 30$ au were analyzed. Such an approach seems more justified in the case of studying dynamical mechanisms other than gravitational perturbations of known planets (for orbits with perihelions near the orbit of Neptune, perturbations from this planet can be significant). The orbits presented on the website of the Minor Planet Center on January 28, 2019 (https://www.minorplanetcenter.net/iau/lists/Centaurs.html) were used, for objects observed in at least two oppositions.

Figure 1 presents the distribution of longitudes of perihelion and semimajor axes of distant trans-Neptunian objects. As already noted in (Batygin, Brown, 2016; Batygin et al., 2019), there are two groups of objects. The first group is concentrated approximately at 60–70°, and the longitudes of perihelion for the second group differ by approximately 180°.

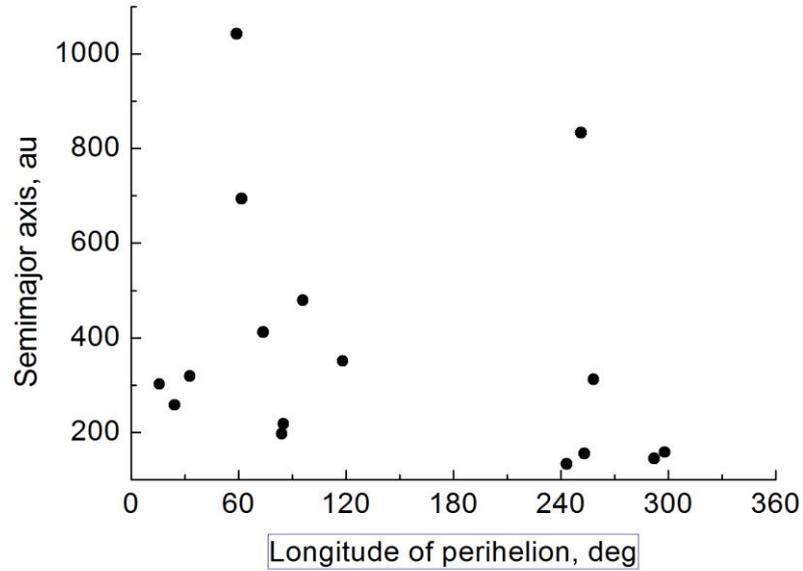

**Fig. 1.** Distribution of longitudes of perihelion and semimajor axes for observed multiple-opposition trans-Neptunian objects with $q > 40$ au and $a > 100$ au.

Figure 2 shows a large scatter of longitudes of ascending nodes for the same objects. An interesting feature is the lack of distant trans-Neptunian objects with $a < 1000$ au in the range of longitudes of ascending nodes (240, 360) degrees, but the statistical data are still insufficient for a confident statement about the reliability of this feature.

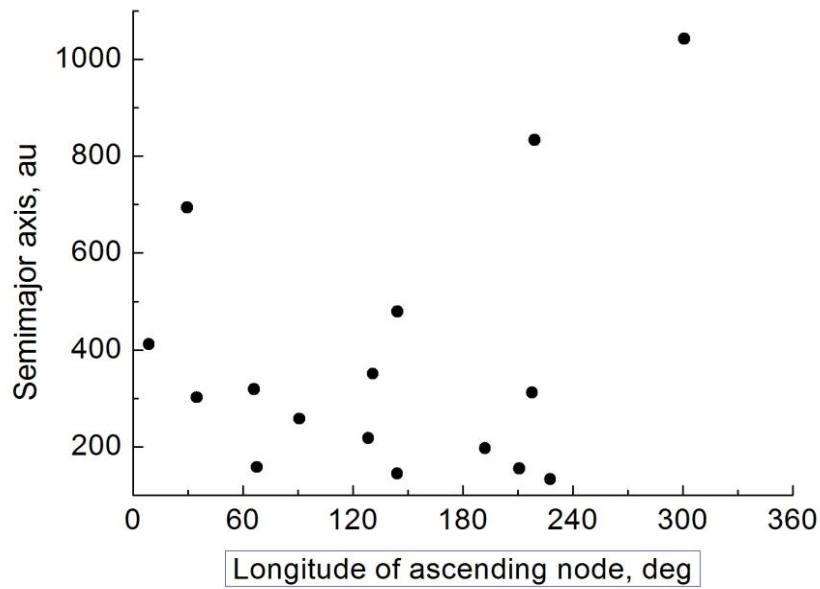

**Fig. 2.** Distribution of longitudes of ascending nodes and semimajor axes for observed multiple-opposition trans-Neptunian objects with $q > 40$ au and $a > 100$ au.

Figure 3 demonstrates the distribution of arguments of perihelion for distant trans-Neptunian objects. At the initial stage of the discussion of the possible existence of the ninth planet in the work (Trujillo, Sheppard, 2014), it was suggested that a concentration of arguments of perihelion near the value $\omega = 0°$ is the case; however, further studies did not confirm this assumption. As can be seen from Fig. 3, there is no visible grouping of arguments of perihelion for distant trans-Neptunian objects.

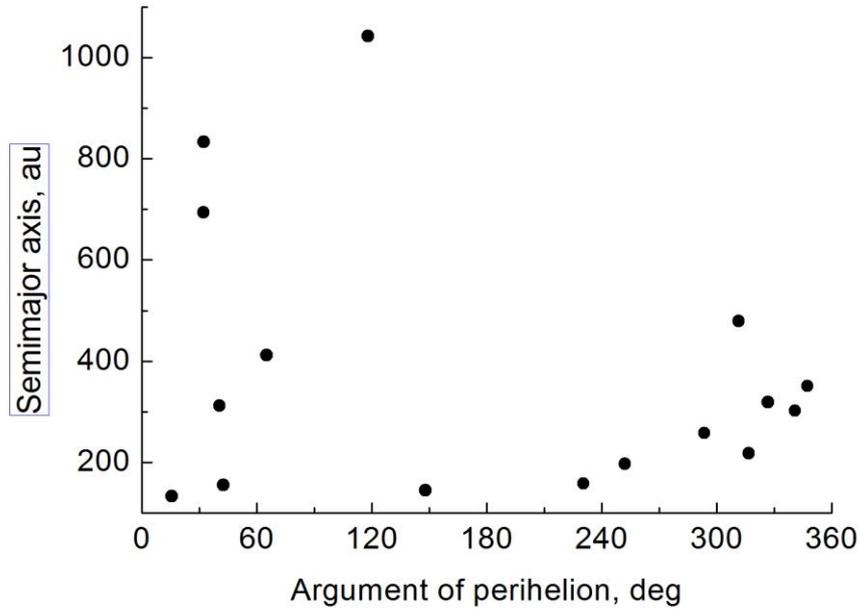

**Fig. 3.** Distribution of arguments of perihelion and semimajor axes for observed multiple-opposition trans-Neptunian objects with $q > 40$ au and $a > 100$ au.

Thus, only for the distribution of longitudes of perihelion the grouping of distant trans-Neptunian objects near certain values is clearly seen in modern observational data. There are two groups of objects concentrated near the longitudes of perihelion that differ by approximately 180°. According to a statistical analysis in (Brown, Batygin, 2019), the observed longitudes of perihelion are not a sample from a uniform distribution at a significance level of 0.04.

Another feature of the distribution of distant trans-Neptunian objects is shown in Figure 4. The lack of distant trans-Neptunian objects with $a < 1000$ au is seen in the interval of perihelion distances (50, 75) au, although the statistics are not yet sufficient for a confident statement about the reliability of this feature.

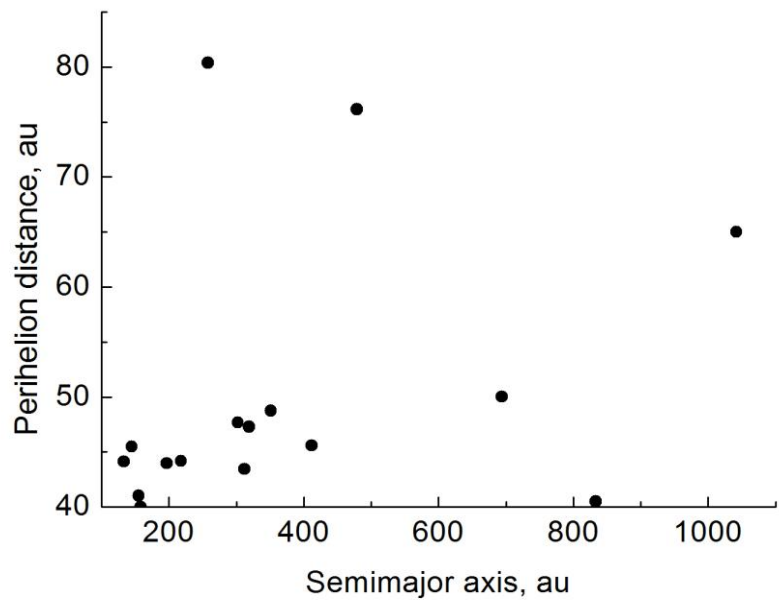

**Fig. 4.** Distribution of semimajor axes and perihelion distances for observed multiple-opposition trans-Neptunian objects with $q > 40$ au and $a > 100$ au.

MODEL

We consider a typical example of the dynamical evolution of two interacting giant clumps in the outer part of the protoplanetary disk, shown in Fig. 17 in (Vorobyov, Elbakyan, 2018). An outer object with a lower mass moves in an orbit with eccentricity of ~0.5. The inner object moves in an almost circular orbit with $a \sim 100$ au. After a certain time period there is a close encounter of these objects. As a result of mutual perturbations, the objects enter orbits with large eccentricities. The outer object begins to move along the hyperbola and is ejected from the system of the star under consideration. The inner object enters an orbit with $q \sim 30$ au, and its orbit is quickly rounded off at such a distance due to interaction with the protoplanetary gas disk.

We are trying to find out how the model of giant gaseous clumps under consideration is consistent with the observed distribution of distant trans-Neptunian objects, and thus, are there any grounds for considering it with respect to the Solar system. We assume that the population of small bodies existed at the stage of migration of gaseous clumps. In our model, small bodies initially moved in almost circular orbits with small inclinations to the ecliptic. In the calculations (Vorobyov, Elbakyan, 2018), the lifetime of giant clumps in the protoplanetary disk does not exceed 0.5 Myr. Although the question of the formation of small bodies in such a short time period is open, there are many works that consider the very rapid formation of small bodies in the outer Solar system. For example, the authors of the work (Wahlberg Jansson, Johansen, 2014) claim that in their numerical experiments planetesimals of various sizes can be formed at 40 au by the streaming instability: more than ~100 km in ~25 years, 10 km in several hundred years, 1 km in several thousand years.

The dynamical processes of evolution of giant clumps are very complex, while the masses of the clumps are variable, and their orbits are disturbed. Now it is impossible to establish all the details of the dynamics of giant clumps. Therefore, in this paper, we study successively the various simplest elements of this process, examining objects with constant masses moving in Keplerian orbits: 1) the motion of an outer object in the orbit with eccentricity $e_c \sim 0.5$; 2) the motion of an inner object in an almost circular orbit; and 3) the motion of an inner object from aphelion to perihelion with $q_c \sim 30$ au. The effect of each dynamical episode on the distribution of the orbits of small bodies located in the region of the gravitational influence of giant clumps at time intervals not exceeding several tens of thousands years according to (Vorobyov and Elbakyan, 2018) is studied.

In each case, the motion of a large number of particles experiencing gravitational action from giant clumps was considered. Integration of the equations of motion was performed using a symplectic integrator (Emel'yanenko, 2007). Integration for a given particle was terminated if $a > 1000$ au or $q < 0.2$ au, and also in the event of a collision of a particle with a giant clump. At the initial moment of time, the semimajor axes of the particles were randomly distributed according to the law $a^{-0.5}$ in a certain interval, and the eccentricities $e$ and the orbital inclinations $i$ of the particles were uniformly distributed random variables with $e < 0.001$ and $i < 0.5°$.

MOTION OF THE OUTER OBJECT IN A HIGH-ECCENTRICITY ORBIT

Let us consider a model in which an object with mass $m_c = 10 m_J$, radius $R_c = 4$ au moves in an orbit with the semimajor axis $a_c = 600$ au, the eccentricity $e_c = 0.5$, the inclination $i_c = 20°$, the longitude of perihelion $\pi_c = 61°$, and the longitude of ascending node $\Omega_c = 91°$. The choice of angular orbital elements is based on the same arguments that were considered in (Batygin, Brown, 2016; Batygin, Morbidelli, 2017) to search for a hypothetical ninth planet. The initial values of semimajor axes of 5000 particles are distributed in the interval (50, 1000) au.

Figure 5 shows the distribution of longitudes of perihelion and semimajor axes of particles with $40 < q < 80$ au and $a > 100$ au after 40 000 years. In the distribution of the modeled particles there are concentrations of longitudes of perihelion near $\pi_c$ and $\pi_c + 180°$, similar to the distribution of the observed distant trans-Neptunian objects.

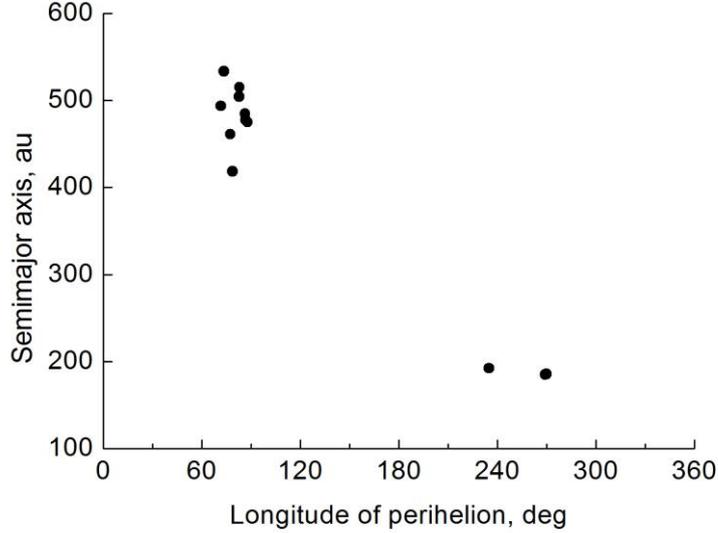

**Fig. 5.** Distribution of longitudes of perihelion and semimajor axes for distant particles after 40 000 years of evolution under the action of gravitational perturbations from a giant clump with $a_c = 600$ au.

We note that the mechanism of the formation of these concentrations in the case under study differs from the mechanism of orbital and secular resonances, considered in (Batygin, Brown, 2016; Batygin, Morbidelli, 2017). In this case, the main factor is the close encounters of particles with giant clumps. It is known (Carusi, Valsecchi, 1980; Emel'yanenko, 1997) that the most efficient changes of orbits take place when tangential approaches occur near the perihelion and aphelion of giant clump orbits, which is manifested in the creation of two maxima in the distribution of particles.

## MOTION OF THE INNER OBJECT IN A CIRCULAR ORBIT

Let us consider a model in which an object with mass $m_c = 25 m_J$, radius $R_c = 12$ au moves in an orbit with the semimajor axis $a_c = 90$ au, the eccentricity $e_c = 0$, and the inclination $i_c = 0°$. The initial values of the semimajor axes of 1000 particles are distributed in the interval (5, 500) au.

Figure 6 demonstrates the distribution of semimajor axes and perihelion distances of particles after 40 000 years. One can see the lack of distant particles in the range of perihelion distances (50, 80) au, which is consistent with observational data for distant trans-Neptunian objects in Fig. 4.

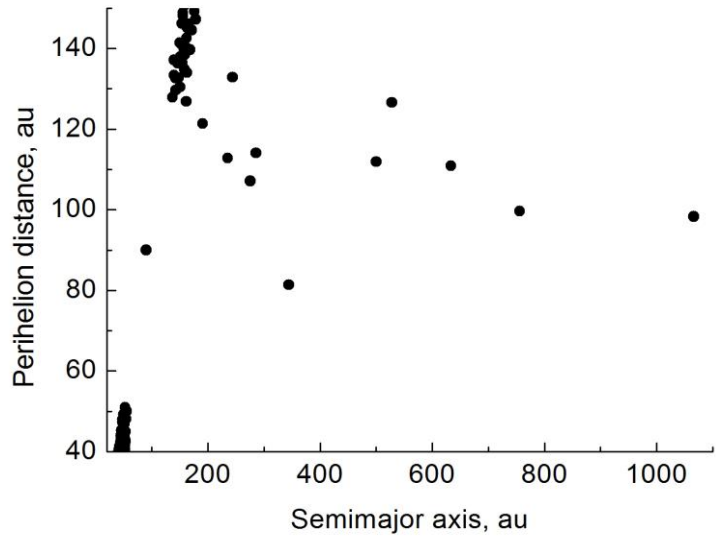

**Fig. 6.** Distribution of semimajor axes and perihelion distances for particles after 40 000 years of evolution under the action of gravitational perturbations from a giant clump with $a_c = 90$ au.

## MIGRATION OF THE INNER OBJECT

Let us now consider the influence of a giant clump on the orbital distribution of small bodies during the transition to the inner region. To do this, we studied a model in which an object with mass $m_c = 10 m_J$, radius $R_c = 5$ au moves in the ecliptic plane from aphelion to perihelion in an orbit with $a_c = 50$ au and $e_c = 0.5$. The initial values of semimajor axes of 1000 particles are distributed in the interval (5, 50) au. Figure 7 shows the distribution of particles after about 180 years, which corresponds to the time interval during the transition of a giant clump from aphelion to perihelion.

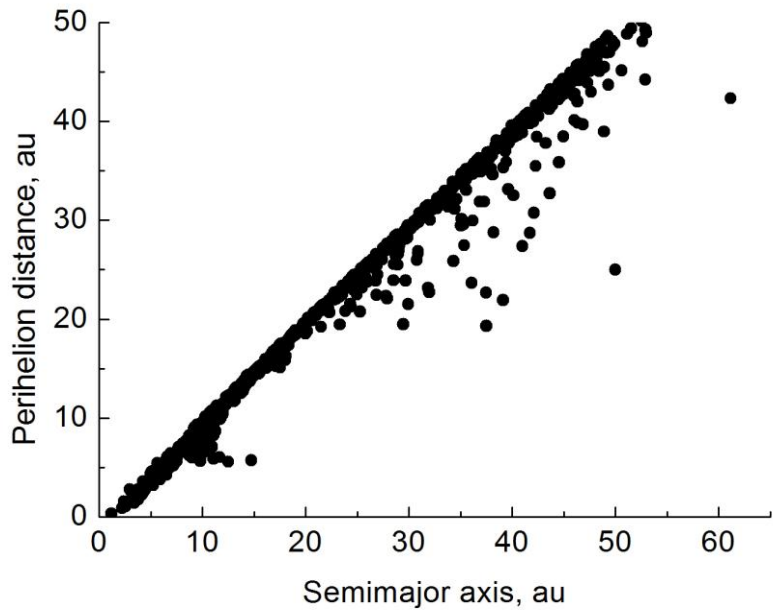

**Fig. 7.** Distribution of semimajor axes and perihelion distances for particles after 180 years of evolution under the action of gravitational perturbations from a giant clump moving from aphelion to perihelion in an orbit with $a_c = 50$ au and $e_c = 0.5$.

It can be seen that for some objects in the Kuiper belt, the eccentricities of the orbits increase significantly. Although modern theories relate the origin of such orbits to the migration of planets in planetesimal disks in the late stages of the Solar system formation (see, e.g., the review (Morbidelli, Nesvorny, 2019)), these calculations show that giant clumps migrating from the outer Solar system could play an important role in increasing the eccentricities of both Kuiper belt objects and more inner planetesimals.

## CONCLUSIONS

New data on the distribution of distant trans-Neptunian objects and on the properties of comets have raised new questions about the dynamical processes in the outer part of the protoplanetary disk in the early stages of the Solar system formation. In this paper, we examined the possible effect of giant gaseous clumps, resulting from gravitational instability and fragmentation of circumstellar disks, on the orbital distribution of the population of small bodies in the outer Solar system. Basically, we studied those features of migration and gravitational interaction of giant clumps that were found in the work (Vorobyov, Elbakyan, 2018).

Our modeling showed that the main features of the distribution of small bodies arising as a result of the gravitational action of giant clumps with masses of the order of $10 m_J$ and more are consistent with the observed distribution of the orbits of distant trans-Neptunian objects. It should be noted that the studied dynamical process associated with a single giant clump is very short in time (no more than several tens of thousands of years). The main factor affecting the orbital distribution of small bodies is close encounters with the giant clump. A significant fraction of small bodies (comets) is very quickly transferred to distant orbits with large eccentricities, which allows them to avoid mutual collisions.

The question of the possible further evolution of giant clumps is open. According to Vorobyov and Elbakyan (2018), giant gaseous clumps exist in the protoplanetary disk for no more than 0.5 Myr, or destroying in the inner part of the disk, or falling into a star, or being thrown into hyperbolic orbits as a result of mutual perturbations. However, there are also very original points of view that these clumps can become embryos of planets, even of terrestrial type (Nayakshin, 2017).

ACKNOWLEDGMENTS

This work was supported by the Russian Science Foundation (project No. 17-12-01441). The calculations were carried out using the MBC-100K supercomputer of the Joint Supercomputer Center of the Russian Academy of Sciences. The author is grateful to I.I.Shevchenko for useful comments.

*Zhu Z., Hartmann L., Nelson R.P., Gammie C.F.* Challenges in forming planets by gravitational instability: disk irradiation and clump migration, accretion, and tidal destruction // Astrophys. J. 2012. V. 746. Article id. 110. 26 p.